\documentstyle[sprocl,psfig]{article}

\newcommand{\gev}{\, {\rm GeV}}

\newcommand{\bea}{\begin{eqnarray}}
\newcommand{\eea}{\end{eqnarray}}
\newcommand{\be}{\begin{equation}}
\newcommand{\ee}{\end{equation}}
\newcommand{\bi}{\begin{itemize}}
\newcommand{\ei}{\end{itemize}}

\newcommand{\beq}{\begin{equation}}
\newcommand{\eeq}{\end{equation}}
\newcommand{\ba}{\begin{array}}
\newcommand{\ea}{\end{array}} 
\newcommand{\beqa}{\begin{eqnarray}}
\newcommand{\eeqa}{\end{eqnarray}}

\newcommand{\no}{\nonumber \\}

\newcommand{\co}{\; \; ,}
\newcommand{\fs}{\; \; .}

\newcommand{\hepph}[1]{\texttt{hep-ph/#1}}

\newcommand{\lbar}{\bar{\ell}}
\newcommand{\Wbar}{\,\overline{\rule[0.75em]{0.9em}{0em}}\hspace{-1em}W}
\newcommand{\Pbar}{\,\overline{\rule[0.75em]{0.5em}{0em}}\hspace{-0.7em}P}

\newcommand{\pbar}{\overline{\rule[0.5em]{0.4em}{0em}}\hspace{-0.5em}p}

\newcommand{\rs}{\langle r^2\rangle^\pi_S}

\def\be{\begin{equation}}
\def\ee{\end{equation}}
\def\bea{\begin{eqnarray}}
\def\eea{\end{eqnarray}}
\begin{document}

\title{RECENT DEVELOPMENTS IN CHIRAL PERTURBATION THEORY}

\author{GILBERTO COLANGELO}

\address{Institut f\"ur Theoretische physik der Universit\"at Z\"urich\\
Winterthurerstr. 190, 8057 Z\"urich\\E-mail: gilberto@physik.unizh.ch} 


\maketitle\abstracts{I briefly review the current status of chiral
perturbation theory (CHPT) in the meson sector. Emphasis is given on the
quest for higher precision. I discuss two examples: one where it is
difficult to make a good prediction ($K_L \to \pi^0 \gamma \gamma$), and
where CHPT, even when pushed to higher orders, cannot yield an increase in
accuracy. The second one is $\pi \pi$ scattering where a very sharp
prediction can be made by combining CHPT at the two--loop level with
dispersion relations.}

\section{Introduction}
Chiral perturbation theory (CHPT) is the low--energy effective theory of
the strong interactions. It yields a systematic expansion of any Green
function in powers of quark masses and momenta, that automatically respects
the Ward identities implied by the chiral symmetry of the strong
interactions. Its present form is due to Weinberg  \cite{weinberg79} and
Gasser and Leutwyler \cite{GL}. They showed the advantages of the
effective--field--theory language over the direct implementation of the
Ward identities as in the current--algebra framework. In particular, what
was known before as a very difficult problem, the calculation of
the unitarity corrections to a current--algebra result, was reduced to a
routine loop calculation in a well--defined framework.  After the
convenient tools of the effective field theory were made available, many
processes have been calculated at the one--loop level: wherever possible,
the comparison to the experimental data has shown a remarkable success of
the method.

In the early nineties several precision experiments on kaons, pions and
etas, were approved, and brought the challenge of increasing the precision
in CHPT calculations. Going beyond one loop became a necessity. The first
complete two--loop calculation concerned the cross section for the
two--photon annihilation into two neutral pions \cite{BGS}.  This beautiful
and difficult calculation opened up the field of two--loop calculations in
CHPT, that was very active during the nineties. In fact if we consider only
the two--light--flavour sector, all the phenomenologically relevant
calculations have already been completed, whereas in the $SU(3)$ framework,
they have started only in more recent years \cite{GK,ABT}. Moreover, in the
purely strong sector, the Lagrangian at order $p^6$ and the complete
divergence structure have been recently calculated \cite{lagrp6}.

In an effective theory it is not obvious that a higher loop calculation may
lead to higher precision: new constants come into play at each order, and
since they are a priori unknown, they may induce an uncertainty which is as
large as the effect that one is calculating. The quest for higher precision
within an effective field theory is a particularly nontrivial one. In what
follows I will discuss in some detail why it is nontrivial, and how, in
some special cases, one can nevertheless succeed in making very accurate
predictions.

\section{The Lagrangian of chiral perturbation theory}
At lowest order the Lagrangian of CHPT is remarkably simple\footnote{For
  simplicity's sake I do not write down explicitly the 27plet, nor the the
  $CP$--odd part of the weak Lagrangian.}:
\begin{eqnarray}
{\cal L}^{(2)} &=&{F^2\over 4} \langle D_\mu U D^\mu U^\dagger +
U\chi^\dagger+\chi U^\dagger \rangle + C \langle QU Q U^\dagger \rangle
\nonumber \\ 
&+& c_2 \langle \lambda_6 D_\mu U^\dagger D^\mu U\rangle + c_3
  {(27_L,1_R)} \; \; ,
%\nonumber \\
%&+& c_2^- \langle \lambda_7 D_\mu U^\dagger D^\mu U\rangle + c_3^-
%  {(27_L,1_R)} \; \; ,
\end{eqnarray}
where
\begin{eqnarray}
D_\mu U&=&\partial_\mu U -ir_\mu U+iUl_\mu \; \; \; \;  \chi =2B(M+s+ip)
\nonumber \\  
M&=&\mbox{diag}(m_u,m_d,m_s) \; \; \; \ Q=e\cdot \mbox{diag}(2/3,-1/3,-1/3)
\; \; , 
\end{eqnarray}
where $U$ is the usual exponential of the matrix containing the octet of
the Goldstone bosons, while $r_\mu, \; l_\mu, \; s$ and $p$ are
matrices of external fields. $\langle A \rangle$ stands for the trace of
$A$. The simplicity of this Lagrangian is particularly remarkable in view
of the variety of phenomena that it describes: the first two terms describe
the strong interactions of the octet of the pseudoscalars between
themselves and with external sources. The third, proportional to $C$
describes the electromagnetic effects due to the exchanges of one virtual
photon, while the remaining terms account for the effects generated by
virtual exchanges of a $W$ or a $Z$ between quarks.

The five constants which appear in the Lagrangian are not fixed by the
symmetry but are easily determined in the comparison to the simplest of the
observables:
\begin{eqnarray}
F_\pi&=&F+O(m) \; \; , \; \; \; \; M_{\pi^0}^2=B(m_u+m_d) \; \; ,
\nonumber \\
M_{\pi^+}^2-M_{\pi^0}^2&=&2e^2 {C\over F^2} \; \; , \; \; \;
%\nonumber \\
A^{(1/2)}_{K\to \pi \pi} \sim c_2-{2 \over3} c_3 \; , \; \; A^{(3/2)}_{K\to
  \pi \pi} \sim c_3 \; . 
\end{eqnarray}
Having fixed these constants, we can now calculate any other new
observable, and make a prediction. For example the $\pi \pi$ scattering
amplitude and the $K_{l3}$ form factors are given by:
\begin{equation}
A(s,t,u)= {s-M_\pi^2 \over F_\pi^2} \Rightarrow a_0^0={7 M_\pi^2
  \over 32 \pi F_\pi^2} = 0.16 \; , 
%\; a_0^2={-M_\pi^2 \over 16 \pi  F_\pi^2} = -0.045\; \; , 
\end{equation}
%which yields for the $S$--wave scattering lengths:
%\be
%a_0^0={7 M_\pi^2 \over 32 \pi F_\pi^2} = 0.16 \; \; , \; \; \; \;
%\ee
\begin{eqnarray}
\langle \pi^-(p)|V_\mu | K^0(k) \rangle&=&(k+p)_\mu f_+(t)+(k-p)_\mu
f_-(t)\; \; , \; \; \; t=(k-p)^2 \; \; \nonumber \\
f_+(t) &=& 1 \; \; , \; \; \;
f_-(t) = 0 \; \; .
\end{eqnarray}

\subsection{The Lagrangian at next--to--leading order and beyond: strong
  sector}
In the strong sector the situation is the following:
\begin{eqnarray*}
O(p^2)&\Rightarrow&{F^2\over 4} \langle D_\mu U D^\mu U^\dagger +
U\chi^\dagger+\chi U^\dagger \rangle \; \; , \\ 
O(p^4)&\Rightarrow& \mbox{10 new LEC, Gasser and Leutwyler \cite{GL} (84)},
\; \; 
L_i\; \; ,  \\
O(p^6)&\Rightarrow& \mbox{96 new LEC, Bijnens, G.C., Ecker \cite{lagrp6}
  (99)}, \; \; C_i \; \; . 
\end{eqnarray*}
For all the ten $L_i$'s we have either a good phenomenological
determination, or a sound theoretical estimate. In order to get a
prediction at order $p^4$ in this sector one simply needs to do the
one--loop calculation, remove the divergences with the help of the
counterterms, and insert the numerical values of the finite parts from the
most recent estimates. At this level the situation is almost as good as at
order $p^2$. This is best seen with a couple of examples, like the $\pi
\pi$ scattering lengths and the $K_{e3}$ form factors, that we have seen
before at leading order. The next--to--leading order corrections
are \cite{GL,GLFF}
\begin{eqnarray}
\label{a00_1loop}
a_0^0\!&=&\!{7 M_\pi^2 \over 32 \pi F_\pi^2}\left[1+{M_\pi^2 \over 3} \langle
  r^2 \rangle_S^\pi+ {200\pi F_\pi^2 M_\pi^2 \over 7}(a_2^0+2 a_2^2)
+{ M_\pi^2 \over 16 \pi^2 F_\pi^2} \Delta_0^0 \right] \no
&=& 0.16\cdot\left[ 1.25 \right] =0.20 \; \; ,
\end{eqnarray}
\begin{eqnarray}
\label{Ke3_1loop}
f_+(t) &=& f_+(0)\left(1+\lambda_+ {t \over M_\pi^2} \right) \; , \; 
\lambda_+={M_\pi^2 \over 6}\langle r \rangle^\pi_V+\Delta_+ = 0.031 \; \; .
\end{eqnarray}
The formulae above neatly illustrate the beauty of the method: the
$S$--wave scattering lengths and the $K_{e3}$ form factors are expressible
in terms of other observables (like the scalar and vector radii, $\langle
r^2 \rangle_S^\pi$ and $\langle r \rangle^\pi_V$, respectively, the two
$D$--wave scattering lengths, $a_2^{0,2}$), plus other small corrections
(indicated by the $\Delta$ symbols). The relation between these observables
is exact up to order $p^4$, and will get corrections if we want to increase
the precision of the calculation. 
By inserting the most recent values for the observables on the right--hand
sides, one gets the numerical predictions given above. We will discuss the
situation for the $\pi \pi$ scattering lengths in detail below. As for the
$K_{e3}$ form factors, the slope $\lambda_+$ is quite well known, and the
PDG \cite{PDG} reports a number which is in excellent agreement with the
one loop calculation above: $\lambda_+=0.0286\pm0.0022$.

If we move up to order $p^6$ the situation worsens considerably, and it
becomes impossible to write down expressions as nice as those in
Eq. (\ref{a00_1loop}\ref{Ke3_1loop}). We have to give expressions in terms of
the new low--energy constants $C_i$'s, and since we know practically
nothing about them, it is an obvious question whether these calculations
represent an improvement at all in the final numerical predictions. 
Before discussing the matter in more detail in the following sections, we
can already anticipate an answer at the intuitive level: The low--energy
constants parametrise our ignorance of the dynamics which is not explicitly
present in the formalism. They contain contributions from all the physics
above the scale of the pseudoscalar mesons -- the most prominent effect is
the one due to the lowest--lying resonances, as it has been explicitly
shown \cite{EGPR} at order $p^4$. Whenever a resonance or any other
``high--energy'' phenomenon is expected to influence substantially a
low--energy quantity, one should expect a very important role of the
low--energy constants, and a serious difficulty in making a precise
prediction. On the other hand, observables dominated by the dynamics of the
pseudoscalar, can be very well calculated within CHPT, and it makes good
sense to push the calculations to order $p^6$, as one can expect the
contribution of the low--energy constants to be suppressed. Explicit
examples of both situations are given below.

\subsection{The Lagrangian at next--to--leading order and beyond: weak and
  electromagnetic sectors}
Both in the electromagnetic and weak sector, the situation becomes
unpleasant already at next--to--leading order, where the number of
constants immediately gets too large in comparison to the available
experimental input:
\begin{eqnarray*}
\mbox{E.m. sector:}&&\\
O(p^2)&\Rightarrow&C \langle QU Q U^\dagger \rangle\; \; , \\
O(p^4)&\Rightarrow& \mbox{21 new LEC, Urech, Neufeld and
  Rupertsberger  \cite{urech}
  (93)}, \; \; K_i \; \; ,  \\
\mbox{Weak sector:}&&\\
O(p^2)&\Rightarrow& c_2 \langle \lambda_6 D_\mu U^\dagger D^\mu U\rangle \\
O(p^4)&\Rightarrow& \mbox{37 new LEC, Kambor, Missimer, Wyler, Ecker
   \cite{weak_lagr} (91)},
\; \; N_i \; \; .
\end{eqnarray*}
In both cases, given the practical unfeasibility of a determination of all
the available constants, the typical strategy has been to focus on a few
selected processes where the same constants occur. One can then exploit the
experimental information on some processes to make predictions in other.
There are plenty of examples in the literature. A fairly comprehensive
review of the situation in the weak sector (especially for the radiative
decays), including a list of the new experimental measurements that are
possible at a $\Phi$ factory can be found in Ref.  \cite{dafne_hdbk_dein}.

%Neither for the electromagnetic, nor for the weak sector, has anybody tried
%to write down the Lagrangian at order $p^6$.

\section{Analysis of $K \to \pi \gamma \gamma$ up to order $p^6$.}

Assuming $CP$ conservation the $A(K_L\to \pi^0 \gamma \gamma)$ is
determined by two invariant amplitudes, $A(s,\nu)$ and $B(s,\nu)$,
$s=(q_1+q_2)^2, \nu=p_K\cdot(q_1-q_2)$, where $q_{1,2}$ are the momenta of
the two photons, and $p_K$ that of the kaon.  At order $p^2$: $A=B=0$.  At
order $p^4$ \cite{Kpigg_p4}: $A = 4/ s (s-M_\pi^2) F(s / M_\pi^2) + \ldots$,
and $B=0$, where $F(x)$ is a loop function generated by the $\pi \pi$
intermediate state in the $s$ channel, that represents the dominant effect
at this order, and the ellipsis stands for other less important
contributions. Although the shape of the spectrum was nicely confirmed by
the experiment \cite{E731,NA31}, the branching ratio was a factor three too
small:
\be
BR = \left\{
\begin{array}{lr} (1.7\pm 0.3) \times 10^{-6} &(\mbox{NA31}) \\
(1.86 \pm 0.60 \pm 0.60)\times 10^{-6}        &(\mbox{E731}) \\
0.67 \times 10^{-6} &~~~~~~O(p^4) \; \; ,
\end{array} \right.
\ee therefore requiring large $O(p^6)$ corrections.  The calculations at
order $p^6$ \cite{Kpigg_p6} have considered only the (possibly dominant)
corrections to the pion loops, and added to this a polynomial contribution:
\begin{eqnarray*}
A&=&{4 \over s} (s-M_\pi^2) \tilde F\left({s \over M_\pi^2} \right) +
4 a_V\frac{3M_K^2-s-M_\pi^2}{M_K^2} + \ldots \\
B&=&\tilde G\left({s \over M_\pi^2} \right) -8 a_V + \ldots \; \; ,
\end{eqnarray*}
where $\tilde F(x)$ and $\tilde G(x)$ also come from the $\pi \pi$
intermediate state in the $s$ channel. To get into agreement with the
experiment one needed to have a large and negative $a_V$: $BR=0.83 \times
10^{-6}$ with $a_V=0$ and $BR=1.60 \times 10^{-6}$ with $a_V=-0.9$.  Also
for the spectrum, unitarity corrections alone were not sufficient (and
actually worsened the comparison), while an improved agreement with the
data is obtained only with $a_V\sim-0.9$.

The outcome of this $O(p^6)$ analysis is therefore a clear need for a very
large contribution from the polynomial part. Do we have a dynamical
explanation for such a large constant? I find it instructive to go back by
seven years and see how the situation appeared then: Cohen, Ecker and Pich in
Ref. \cite{Kpigg_p4} described the situation as follows: ``Several model
estimates of $a_V$ have been made in the literature.  A fair summary of
those attempt is that we know neither the sign nor the magnitude of
$a_V$.''  More recently, however, D'Ambrosio and Portol\'es \cite{DAP97}
have built a Vector Resonance Model that does indeed get the right sign and
size for this constant: $a_V^{DP} \simeq -0.72$. This number is now in
amazing agreement with the one extracted from a fit to the most recent data
\cite{KTeV}.

Although D'Ambrosio and Portol\'es estimate of $a_V$ was only a
postdiction, it is reassuring to have an understanding of the size of this
constant. In fact it is not the only case where one can find this relative
size between the various contributions in the chiral expansion. A
well--known analogous example in the strong sector is the vector form
factor.  Its Taylor expansion around $s=0$ is usually defined as $ F_V(s) =
1 + 1/6 \langle r^2 \rangle^\pi_V s + c_V^\pi s^2 +O(s^3)$.  $c_V^\pi$
vanishes at order $p^2$, and can be predicted with no parameters at order
$p^4$. Nowadays it is known up to order $p^6$  \cite{BCT}: 
\bea c_V^\pi&=&
\frac{1}{960 \pi^2M_\pi^2 F_\pi^2} + \frac{1}{(16 \pi^2 F_\pi^2)^2} \left[
\mbox{``}\ln {M_\pi^2 \over \mu^2}\mbox{''}+ \mbox{``}l_i^r\mbox{''}
\right] + \frac{r^r_{V2}}{F_\pi^4} \\ &=&(0.62 + 1.96 + 1.3\cdot 10^{-4}
r^r_{V2}(M_\rho)) \gev^{-4} = 5.4 \gev^{-4}\;, \nonumber 
\eea 
where the latter value is determined experimentally. Here also: 
i) the order $p^4$ parameter--free prediction fails badly;
 ii) there are large $O(p^6)$ unitarity corrections;
 iii) but even larger $O(p^6)$ polynomial contributions,
coming from the well--known $\rho$ resonance.

\section{$\pi \pi$ scattering at order $p^6$}
This is the ``golden reaction'' for Chiral Perturbation Theory: at
threshold the naive expansion parameter is $M_\pi^2/1 \gev^2 \sim 0.02$,
and already a tree level calculation \cite{weinberg66} should be rather
accurate, in principle. However, this rule of thumb is quite misleading
here, as it is shown by the fact that both the one--loop \cite{GL83} (see
Eq. (\ref{a00_1loop})) and the two--loop \cite{BCEGS} calculations produced
substantial corrections. The violation of the rule of thumb has a well
known origin, and is due to the presence of chiral logarithms $L=
M_\pi^2/(4 \pi F_\pi)^2 \ln M^2/\mu^2$, which, for $\mu \sim 1$ GeV change
the expansion parameter by a factor four. If we look at the $I=0$ $S$--wave
scattering lengths, e.g., a large coefficient in front of the single (at
one loop) and double (at two loops) chiral logarithms is the main source of
the large correction \cite{GL83,GC95}: 
\be
a^0_0 = \frac{7 M_\pi^2}{32 \pi F_\pi^2} \left\{1 -
  \frac{9}{2}L  + \frac{857}{42} L^2 + \ldots \right\} =0.217
\label{eq:a0_2loop}
\ee
The final number is the estimate given by the authors of the complete
two--loop calculation \cite{BCEGS}, and was given without an estimate of
the uncertainties, that required considerably more work. A better numerical
prediction with an accurate estimate of the uncertainties is now
available \cite{CGL}, but before describing how this was obtained, it is
useful to analyse in more detail the number in Eq. (\ref{eq:a0_2loop}).

In particular it is important to check the role of the new (and largely
unknown) $O(p^6)$ constants. In this case their influence is rather small:
to obtain the number in Eq. (\ref{eq:a0_2loop}) they were estimated using
resonance saturation  \cite{BCEGS}. However, if one puts them to zero, the
change in $a_0^0$ is less than 1\%. For the analytic part
of the chiral expansion the rule of thumb works quite well.
On the other hand it is more worrisome the effect of the uncertainties in
the $O(p^4)$ constants, supposedly known quite well: changing the constants
within their error bars changes the value of $a_0^0$ typically by 0.01:
The two--loop contribution can be completely overshadowed by
the uncertainty coming from the $O(p^4)$ constants -- to improve the
prediction one has to reduce drastically this uncertainty. This improvement
in precision has been obtained by the use of a different method to
determine the low--energy constants which we are now going to describe.

\subsection{Dispersive representation of the scattering amplitude}

As shown by Roy  \cite{Roy}, the fixed-$t$ dispersion relations can be
written in such a form that they express the $\pi\pi$ scattering amplitude
in terms of the imaginary parts in the physical region of the $s$-channel.
The resulting representation for $A(s,t,u)$ contains two subtraction
constants, which may be identified with the scattering lengths $a^0_0$ and
$a^2_0$. Unitarity converts this representation into a set of coupled
integral equations, which was recently examined in detail  \cite{ACGL}. The
upshot of that analysis is that $a_0^0$ and $a_0^2$ are the essential low
energy parameters: Once these are known, the available experimental data
determine the behaviour of the $\pi\pi$ scattering amplitude at low
energies to within remarkably small uncertainties.

The branch cut generated by the imaginary parts of the partial waves with 
$\ell\geq2$ starts manifesting itself only at $O(p^8)$. Accordingly, we may
expand the corresponding contributions to the dispersion integrals
into a Taylor series of the momenta. The singularities due to the imaginary
parts of the $S$- and $P$-waves, on the other hand, show up
already at $O(p^4)$ -- these cannot be replaced
by a polynomial. We subtract the corresponding dispersion 
integrals as many times as is needed in the chiral representation, for
later convenience:
\bea\label{dispWbar}\Wbar^I(s)& \! \!=& \! \! 
\frac{s^{4-\epsilon_I}}{\pi}\int_{4M_\pi^2}^{\infty}
ds'\;\frac{\mbox{Im}\,t^I(s')}
{s^{\prime\,4-\epsilon_I}(s-4M_\pi^2)^{\epsilon_I}(s'-s )}\; \; ,\eea
where $\{\epsilon_0,\epsilon_1,\epsilon_2\}=\{0,1,0\}$.
Since all other contributions can be replaced by a polynomial, the
phenomenological amplitude takes the form
\bea A(s,t,u)& \! \!=& \! \! 16\pi a_0^2+
\frac{4\pi }{3M_\pi^2}\,(2a_0^0-5a_0^2)\,s+
\Pbar(s,t,u) 
%\\& \! \!& \! \! 
+32\pi\left\{\mbox{$\frac{1}{3}$}
\Wbar^0(s) -\mbox{$\frac{1}{3}$} \Wbar^2(s) 
\right. \no & \! \! + & \! \! \left.
\mbox{$\frac{3}{2}$}(s-u)\Wbar^1(t) +\mbox{$\frac{3}{2}$}(s-t)\Wbar^1(u)
%\right. \no
%& \! \!& \! \! 
+\mbox{$\frac{1}{2}$}\Wbar^2(t)+
\mbox{$\frac{1}{2}$}\Wbar^2(u)
 \right\}+O(p^8)\,.\nonumber\eea
We have explicitly displayed the contributions from the subtraction 
constants $a_0^0$ and $a_0^2$.
The term $\Pbar(s,t,u)$ is a crossing symmetric polynomial
\bea \Pbar(s,t,u)& \! \!=& \! \! \pbar_1+\pbar_2\,s+\pbar_3\,s^2+\pbar_4\,(t-u)^2+
\pbar_5\,s^3+\pbar_6\,s(t-u)^2\fs\eea
Its coefficients can be expressed in terms of integrals over the imaginary
parts of the partial waves. In the following, the essential point is that the
coefficients $\pbar_1,\ldots\,,\,\pbar_6$ can be determined
phenomenologically \cite{ACGL}. 

In the chiral representation we have, similarly
\bea A(s,t,u)
& \! \!=& \! \! C(s,t,u)+32\pi\left\{\mbox{$\frac{1}{3}$}\,U^0(s)+
\mbox{$\frac{3}{2}$}\,(s-u)\,U^1(t)
+\mbox{$\frac{3}{2}$}\,(s-t)\,U^1(u)\right.\no
& \! \!& \! \! \left.+\mbox{$\frac{1}{2}$}\,U^2(t)+\mbox{$\frac{1}{2}$}\,U^2(u)
-\mbox{$\frac{1}{3}$}\, U^2(s) \right\}+ O(p^8)\co \nonumber \eea
where $C(s,t,u)$ is a crossing symmetric polynomial,
\[
C(s,t,u)=c_1+s\,c_2+s^2\,c_3
+(t-u)^2\,c_4+s^3\,c_5
+s\,(t-u)^2\,c_6 \; \; , \]
whose coefficients are given in terms of the low--energy constants
\cite{BCEGS}. 
The functions $U^0(s)$, $U^1(s)$ and $U^2(s)$ describe the 
``unitarity corrections'' \cite{KMSF,BCEGS}  associated with $s$-channel
isospin $I=0,1,2$, respectively, and are the chiral expansion (up to and
including order $p^6$) of the $\Wbar^I(s)$, Eq. (\ref{dispWbar}).

\subsection{Matching the two representations}
In their common domain of validity, the dispersive and the chiral
representations of the scattering amplitude agree,
provided the parameters occurring therein are properly matched.
Since the differences between the functions $\Wbar^I(s)$ and $U^I(s)$
are beyond the accuracy of the two-loop representation   \cite{KMSF},
the two descriptions agree if and only if the polynomial parts
do,
\bea 
\label{mc}
C(s,t,u)=16\pi a_0^2+
\frac{4\pi }{3M_\pi^2}\,(2a_0^0-5a_0^2)\,s+\Pbar(s,t,u)+O(p^8)\fs
\eea 
Since the main uncertainties in the coefficients of the polynomial
$\Pbar(s,t,u)$ arise from their sensitivity to the scattering lengths
$a^0_0$, $a^2_0$, the above relations essentially determine the
coefficients $c_1,\ldots,\,c_6$ in terms of these two observables.  
On the other hand, these coefficients in the chiral representation are not
completely free, but are given in a power series in the pion mass. For
example the combinations
\bea
C_1 &\equiv & F_\pi^2\left\{c_2+4M_\pi^2(c_3-c_4)\right\}
\,,\hspace{0.5em}C_2 \equiv \frac{F_\pi^2}{M_\pi^2}
\left\{-c_1+4M_\pi^4(c_3-c_4)\right\}\, ,\nonumber \eea
satisfy the following low--energy theorems
\[
C_1 = 1+
\frac{M_\pi^2}{3}\,\rs+\frac{23\,\xi}{420}+O(\xi^2)\co \; \;
C_2 =
1+\frac{M_\pi^2}{3}\,\rs+ \frac{\xi}{2}\left\{\lbar_3-
\frac{17}{21}\right\} + O(\xi^2)
\]
where $\xi=(M_\pi/ 4\pi F_\pi)^2=0.01445$.
If we use the information on the scalar radius and a rough estimate for the
constant $\ell_3=2.9 \pm  2.4$, then two of the $c_i$'s become known,
and we may use the set of equations (\ref{mc}) to determine the two
scattering lengths, and four of the $c_i$'s.

\begin{figure}[thb]
%\rule{5cm}{0.2mm}\hfill\rule{5cm}{0.2mm}
%\vskip 2.5cm
%\rule{5cm}{0.2mm}\hfill\rule{5cm}{0.2mm}
\centerline{\hbox{\psfig{figure=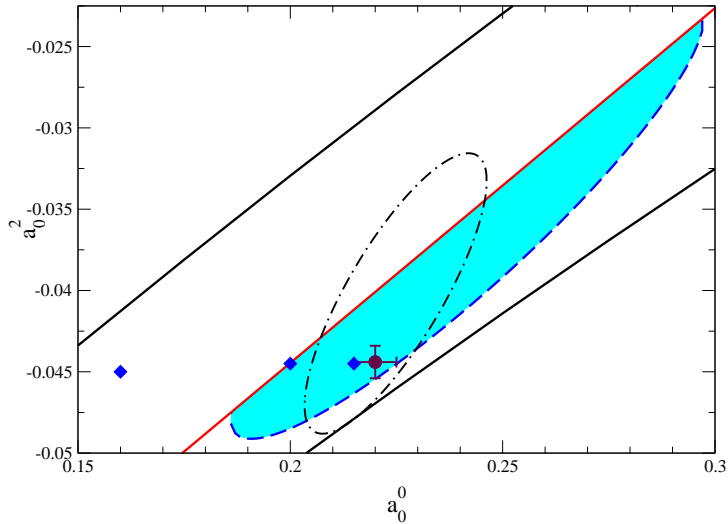,height=8.5cm,angle=-90}}}
\caption{The shaded region represents the intersection of the 
domains allowed by the old data and by the Olsson sum rule, see
\protect \cite{ACGL} for details.
The ellipse indicates the impact of the new, preliminary $K_{e_4}$
data \protect \cite{Pislak}.  The cross shows the result of Ref. \protect
\cite{CGL}. The three diamonds illustrate the 
convergence of the chiral perturbation series at threshold.
The one at the left corresponds to Weinberg's leading order formulae. The
two black solid lines are the boundaries of the universal band
\protect\cite{ACGL} }
\end{figure}

This leads to the following numbers for the chiral expansion of the $I=0$
$S$--wave scattering length: 
\[
a_0^0=0.197\; \; O(p^2) \to 0.2195\; \; O(p^4) \to 0.220\; \; O(p^6) \;
\; ,
\]
which show a remarkably fast and good convergence. The difference of this
procedure with respect to a direct evaluation of the scattering length is
that here the chiral expansion is only applied in the unphysical region,
where it converges best -- the continuation to threshold to
evaluate the scattering length is then provided by the dispersive
representation, that, as shown in \cite{ACGL}, at low energy does not show
significant uncertainties.
On this basis, we can safely conclude that yet higher orders will not
modify the above predictions beyond the uncertainties present in this
matching procedure. Once this are taken into account, the final result for
the $S$--wave scattering lengths is
\bea\label{final result} \begin{array}{ll}
a_0^0= 0.220\pm 0.005\,,& \hspace{2em}
2a_0^0-5a_0^2= 0.663\pm 0.006\,
\end{array}
\eea

The result (\ref{final result}) relies on the
standard picture, according to which the quark condensate represents the
leading order parameter of the spontaneously broken symmetry.  The scenario
investigated in ref.~ \cite{SSF} concerns the possibility
that the Gell-Mann--Oakes--Renner formula fails, the second term in the
expansion $ M_\pi^2= 2\, B\,m \left\{1-\frac{1}{2}\,\xi\,\lbar_3
  +O(\xi^2)\right\}$ being of the same numerical order of magnitude or even
larger than the first. Note that for this to happen, the value of
$|\lbar_3|$ must exceed the estimate we used by more than an order of
magnitude. 
The constraints imposed on $a_0^0$, $a_0^2$ by the available experimental
information are shown in the figure. The ellipse represents the 68 \% 
confidence level contour obtained by combining the new, preliminary 
$K_{e_4}$ data  \cite{Pislak} with earlier experimental results
 \cite{rosselet}. Concerning the value of $a_0^0$, the ellipse corresponds
to the range $0.2<a_0^0<0.25$. 

The figure shows that the values of $a_0^2$ and $a_0^0$ are strongly
correlated. The correlation also manifests itself in the Olsson sum rule
 \cite{Olsson}, which according to ref.~ \cite{ACGL} leads to
$2a_0^0-5a_0^2=0.663 \pm0.021+1.13 \Delta a_0^0 -1.01\Delta a_0^2$, in
perfect agreement with the result in eq.~(\ref{final result}). This
combination, however, is not sensitive to $\ell_3$ -- accurate 
experimental information in the threshold region is needed to perform a
thorough test of the theoretical framework that underlies the
calculation. The forthcoming results from Brookhaven  \cite{Pislak}, CERN
 \cite{DIRAC,cernkl4} and Frascati  \cite{DAFNE} will provide such a test.

\section{Conclusions}
We have reviewed recent developments in CHPT, emphasising in particular the
problems that arise when pushing the chiral expansion to order $p^6$. We
have discussed a few examples that show different physical situations: in
one case the low--energy constants at order $p^6$ have a dominant role,
whereas in the other only a marginal one. In the former case CHPT cannot
provide a solid prediction with a good control of the uncertainties,
whereas in the latter case, it is possible to yield a very precise
prediction. The case of $\pi \pi$ scattering has been described in some
detail, and we have shown that by combining a dispersive representation and
the chiral expansion one can bring the uncertainties down to the few
percent level \cite{CGL}.

\section*{Acknowledgements}
It is a pleasure to thank the organisers for the invitation to such an
interesting conference, and for their perfect organisation. This work is
partly supported by Schweizerische Nationalfonds.

\end{document}